# POINTS - Playful objects for inclusive, personalized movement games

**Georg Regal**
Center for Technology Experience
Austrian Institute of Technology
Giefinggasse 2, Vienna, Austria
georg.regal@ait.ac.at

**David Sellitsch**
Center for Technology Experience
Austrian Institute of Technology
Giefinggasse 2, Vienna, Austria
david.sellitsch@ait.ac.at

**Elke Mattheiss**
Center for Technology Experience
Austrian Institute of Technology
Giefinggasse 2, Vienna, Austria
elke.mattheiss@ait.ac.at

**Manfred Tscheligi**
Center for Technology Experience
Austrian Institute of Technology
Giefinggasse 2, Vienna, Austria
manfred.tscheligi@ait.ac.at
&
University of Salzburg
Center for Human-Computer Interaction
Sigmund Haffner Gasse 18,
Salzburg, Austria
manfred.tscheligi@sbg.ac.at

## Abstract

Promoting exercise and promoting fun in sport and activity is a common goal of schools. However, children and adolescents do not exercise enough, which can favor a number of chronic illnesses. Exercise and sports often require coordination of visual perception and reaction, which is an additional barrier for visually impaired (blind and partially sighted) people. Due to their highly motivating appeal, games promoting physical activity (exertion games) have become increasingly popular. Although accessible exertion games have been developed, they do not consider the different abilities of players. Especially in team sports player roles that consider individual abilities can foster inclusion. To personalize roles and assign certain abilities to players, wearable technology can play an important role. In this position paper we present ideas how digital objects can be used to design exertion games for visually impaired students and we reflect how wearable technology can be used for personalized player roles.

## Author Keywords
Visually Impaired, Exertion Games, Wearables

## ACM Classification Keywords
H.5.m. Information interfaces and presentation (e.g., HCI): Miscellaneous;

> **Bomb Disposal Game**
>
> An example game play that could be built with the proposed playful objects and wearables.
>
> One object acts as a *bomb*. The *bomb* cannot be moved, otherwise it will *explode* (play an explosion sound) and certain other objects (*tools*: a *cutter* and a *screwdriver*) are needed to defuse the bomb.
>
> The players have to find the *screwdriver* and the *cutter* that are placed in the room among other objects (a *lighter*, a *fire extinguisher*). The objects play a distinct sound in order to be traceable by visually impaired students. Students have to bring the *screwdriver* and the *cutter* to the *bomb* for defusing it. When both objects have touched the bomb a victory sound is played. If the wrong *tool* is used the *bomb explodes.*
>
> To personalize the game the tools can only be used by a special player, the bomb disposal specialist.

### Introduction

Promoting exercise and teaching children the fun of sport and activity is a clear goal of schools and the subject of numerous (political) initiatives. Regular physical exercise (in physical education classes and sports weeks) can lay the foundation for a more active and therefore healthier life. Research has shown that "[..] regular physical activity in the primary and secondary prevention of several chronic diseases [..]" ([8] p. 801) and that "physical inactivity is a modifiable risk factor for cardio-vascular disease and a widening variety of other chronic diseases [...] " ([8] p. 801).

Although the positive effects of physical activity are widely known, show that children and adolescents do not exercise sufficiently (cf. [3]). The amount of time children and adolescents spend sitting with entertainment technology is extraordinarily high compared to the amount of physical activity. Moreover visually impaired (blind and partially sighted) people are even less active than people without visually impairments in the same age group [2]. Also the expectation of parents, their children could be physically active, decreases with the level of visual impairment. Similar children's own expectations to be physically active decreases with increases visual impairment (cf. [7]).

Thus strategies to increase physical activity of visually impaired students are needed. Using games to promote physical activities are especially promising, as play is motivating by itself without the need for additional incentives (cf. [6]). The combination of physical activity with interactive computing technology, so called exertion games [5], is particularly interesting.

Exertion games for visually impaired people have been proposed, for example VI Tennis, VI Bowling and Pet-N-Punch [4]. Playing the games resulted in an "increase in energy expenditure for all the games, but the increase could be higher" ([4], p- 169). Thus the authors suggest games that use whole body movements to increase the physical exercise effect.

Therefore we propose exertion games for blind and visually impaired that integrate playful digital objects and wearables to support games that aim for whole body movement. Related work suggested balancing exertion games by adjusting "the power of each player's avatar" ([1], p. 3). Following that suggestion we propose to design digitally augmented team sports with player roles that consider individual abilities of players and foster inclusion by using personalization to make players (dis)abilities an integral part of the game play.

In this position paper we describe first ideas and concepts, how playful objects and wearables can be used to augment sports for visually impaired students.

### Digital Playful Objects for Visually impaired Students

We propose intelligent playful objects that can be used to augment movement based games. We intentionally do not aim to propose one game that is enhanced by digital objects, but propose objects that are generic usable in multiple games. Thus we describe basic features that should be provided by such objects, to support multiple game design strategies.

The objects will consist of sensors to react to the environment and actuators to provide feedback to the player. Keeping the target group of visually impaired people in mind, we propose sound and vibration as feedback modalities. As sensors we propose RFID / NFC reader and accelerometer.

Sound and vibration feedback could be used to find and identify objects (e.g. the "screwdriver" makes a characteristic sound) as well for communicating the status of the object (e.g. play an explosion sound) or the game play (e.g. play an victory sound). Sound has the benefit that it makes objects traceable and identifiable from a distance. Also more than one player can recognize the status of an object. Vibration on the other hand has the benefit that it can communicate a status only to the one player who holds the object, which could be necessary, if the status of an object should be a "secret" to the other players.

RFID / NFC reader and accelerometer are used set the objet to a desired state and to identify other objects. NFC and RFID allow the object to react if other objects are nearby (e.g. set the "bomb" to defuse if the screwdriver and the "cutter" are nearby). The accelerometer can be used to determine the acceleration of the object. For example it could be used to set the object to a certain state if moved, or if not moved (e.g. make a sound while not moved, so visually impaired players can find the object).

### Personalizing Player Roles through Wearables

To personalize the game-play, we propose using wearable devices. According to player's abilities and disabilities certain actions can only be done by certain players. For example objects can only be moved by blind players, to foster including blind players in the gameplay. For example to personalize the bomb disposal game the "screwdriver" and the "cutter" can only be used by the bomb defusing specialist. Moreover a possible further personalization could be that all players can find the objects, but only blind players can move them and have to bring them to the specialist.

We propose two type of wearables for personalizing player roles, "active" and "passive" wearables.

Active wearables contain sensors and actuators that can react to the environment and the player. Such wearables should have the same features as a playful object (sound, vibration, accelerometer and RFID / NFC reader) and could be additionally combined with a heart rate sensor. For example the wearables could play a sound that identifies the player, e.g. as the bomb disposal specialist to allow other players to easily find the player. Also active wearables can react to the body status of a user. For example they can measure how many steps a user can go before playing a sound, in the "bomb disposal game" this could be used to allow players to only make 5 steps with a tool in hand. Afterward the player has to pass the tool to another player. Also heartrate could be measured, e.g. the wearable plays a sound if the heartrate of a user is below a certain threshold.

Passive wearables are mainly used to identify players and player roles. We propose that players wear RFID tags that allow them to be identified by other players and objects. Preferably we will use UHF-RFID tags, as with certain antennas reading passive tags over a distance of 60cm to ~5 meters are possible.

## Conclusion

In this position paper we have presented first ideas how digital objects and wearables can be used to design fully body exertion games for visually impaired students.

By making sports more inclusive and fun for visually impaired students, we aim to increase the level of activity for this target group, as visually impaired people are less active than people without visually impairments [2]. This should result in positive impacts for the health of visually impaired students as regular exercise prevents chronical diseases [8]. Especially we proposed the usage of wearable technology to personalize player roles (cf. [1]) to integrate the different abilities of players in the game design. This should lead to more inclusive exertion games, as certain players (e.g. players who are blind) are needed in a central role for winning a game play.

In future work we will build functional prototypes of the wearables and playful objects and use those porotypes in game design workshops with visually impaired students and teachers to co-design inclusive exertion-games.

## Acknowledgements

We thank the students and teachers of the Austrian Institute for the Blind for their participation.

This work has been partly funded by the Project POINTS – SPA 06/153, a project conducted within the funding program "Sparkling Science", funded by the Austrian Federal Ministry of Education, Science and Research.

## References

1. David Altimira, Florian "Floyd" Mueller, Gun Lee, Jenny Clarke, and Mark Billinghurst. 2014. Towards understanding balancing in exertion games. *Proceedings of the 11th Conference on Advances in Computer Entertainment Technology - ACE '14*: 1–8.

2. Michele Capella-McDonnall. 2007. The Need for Health Promotion for Adults Who Are Visually Impaired. *Journal of Visual Impairment & Blindness* 101, 3: 133–145.

3. Andrew P. Hills, Neil A. King, and Timothy James Armstrong. 2007. The contribution of physical activity and sedentary behaviours to the growth and development of children and adolescents : implications for overweight and obesity. *Sports Medicine* 37, 6: 533–546.

4. Tony Morelli, Eelke Folmer, John T. Foley, and Lauren Lieberman. 2011. Improving the lives of youth with visual impairments through exergames. *Insight: Research and Practice in Visual Impairment and Blindness* 4, 4: 160–170.

5. Florian "Floyd" Mueller, Darren Edge, Frank Vetere, Martin R Gibbs, Stefan Agamanolis, Bert Bongers, and Jennifer G Sheridan. 2011. Designing Sports: A Framework for Exertion Games. In *Proceedings of the SIGCHI Conference on Human Factors in Computing Systems* (CHI '11), 2651–2660.

6. C Scott Rigby. 2015. Gamification and motivation. *The gameful world: Approaches, issues, applications*: 113–138.

7. Moira E Stuart, Lauren Lieberman, and Karen E Hand. 2006. Beliefs about physical activity among children who are visually impaired and their parents. *Journal of Visual Impairments & Blindness* 100, 4.

8. Darren E R Warburton, Crystal Whitney Nicol, and Shannon S D Bredin. 2006. Health benefits of physical activity: the evidence. *CMAJ : Canadian Medical Association journal = journal de l'Association medicale canadienne* 174, 6: 801–9.